\documentclass[twocolumn,english,aps,prd,reprint,floatfix,notitlepage,footinbib,preprintnumbers,superscriptaddress]{revtex4-1}
\pdfoutput=1
\usepackage{lmodern}

\usepackage[T1]{fontenc}
\usepackage[latin9]{inputenc}
\usepackage{geometry}
\usepackage{subfigure,lmodern, amsmath,amssymb, graphicx, pifont, adjustbox, bm, xcolor}
\usepackage{amsfonts}
\usepackage{geometry}
\usepackage{comment}
\usepackage{mathtools}
\usepackage{float}
\usepackage{slashed}
\usepackage{ragged2e}
\usepackage{array}
\usepackage{hhline}

\makeatletter\g@addto@macro\bfseries{\boldmath}\makeatother

\usepackage{stackengine}
\usepackage{esint}
\usepackage[unicode=true,pdfusetitle,
 bookmarks=true,bookmarksnumbered=false,bookmarksopen=false,
 breaklinks=false,pdfborder={0 0 1},backref=false,colorlinks=true]
 {hyperref}
\hypersetup{
 pdfauthor={Clifford Cheung, Ira Z. Rothstein},
 citecolor=black,linkcolor=black,urlcolor=black}

\newcommand{\appendixref}[1]{\hyperref[#1]{appendix~\ref{#1}}}
\def\equationautorefname~#1\null{eq.\,(#1)\null}
\usepackage{breakurl}
\usepackage{breakurl}
\usepackage[hang,flushmargin]{footmisc} 
\allowdisplaybreaks
\makeatletter

\usepackage{etoolbox}
\apptocmd{\thebibliography}{\justifying\setlength{\leftskip}{7.4mm}}{}{} 
 
 \usepackage{relsize}
\usepackage{babel}

\makeatletter
\def\simgt{\mathrel{\lower2.5pt\vbox{\lineskip=0pt\baselineskip=0pt
           \hbox{$>$}\hbox{$\sim$}}}}
\def\simlt{\mathrel{\lower2.5pt\vbox{\lineskip=0pt\baselineskip=0pt
           \hbox{$<$}\hbox{$\sim$}}}}
\makeatother

\usepackage{changepage}

\newcommand{\be}{\begin{equation}}
\newcommand{\ee}{\end{equation}}
\newcommand{\bea}{\begin{eqnarray}}
\newcommand{\eea}{\end{eqnarray}}

\newcommand{\Eq}[1]{Eq.~(\ref{#1})}

\newcommand{\Sec}[1]{Sec.~\ref{#1}}

\newcommand{\App}[1]{App.~\ref{#1}}

\newcommand{\eq}[2]{\be\begin{aligned}#1 \label{#2}\end{aligned}\ee}
\newcolumntype{P}[1]{>{\centering\arraybackslash}p{#1}}

	\definecolor{dartmouthgreen}{rgb}{0.05, 0.5, 0.06}

\newcommand{\CC}[1]{\textbf{\color{red} [CC:] #1}}

\begin{document}

%\title{Is Field Space Unbounded?}

%\title{What's at the End of Field Space?}

%\title{The End of Field Space as We Know It}

%\title{Is there an End to Field Space?}

%\title{The End of Quantum Fields}

%\title{Is Field Space without Limits}

%\title{Effective Field Theory of the End}

\preprint{CALT-TH 2024-046}

\title{New Physics Hiding at the Ends}

\author{Clifford Cheung}
\affiliation{Walter Burke Institute for Theoretical Physics, California Institute of Technology, Pasadena, CA 91125}
\author{Ira Z.~Rothstein}
\affiliation{Department of Physics, Carnegie Mellon University, Pittsburgh, PA 15213}    
    
\begin{abstract}

\noindent Is field space infinite?  If not, it either loops back on itself or ends altogether.  Periodic boundary conditions are of course familiar, but field space endpoints---which appear in real-world systems---are far less explored.   In this paper we argue that boundaries in field space are generic, radiatively stable structures that allow for new physics at very low scales not ruled out by experiment.  Such boundaries are delocalized in field space from the vacuum, so they can only be accessed by coherent fields or high multiplicity processes, both of which are weakly constrained observationally. 
%The scale  of the physics generating the  field space cut-off can be well below  the naive cut-off of an EFT, 
Low multiplicity interactions do not detect the boundary and instead perceive a ``mirage cutoff'' that is parametrically higher than the true cutoff of the theory.  Hence, field space boundaries are deformations of the standard model that are Lorentz invariant, local, unitary at low energies, and experimentally unconstrained.  We comment on the possibility of field space boundaries on the long-range force carriers and the Higgs, as well as possible implications for the hierarchy problem.

\end{abstract}
\maketitle

\section{Introduction} Decades of null experimental results 
%hint at the sobering possibility that
seemingly imply that
 physics beyond the standard model (SM)
 %has not been observed because it 
 is either {\it i}) too massive to be produced, {\it ii}) too feebly interacting to be detected, or {\it iii}) simply not present.  In this paper we argue for a fourth option: {\it iv}) a theory with kinematically accessible states and sizable couplings could still be invisible to current experiments if its interactions require a parametrically high multiplicity of particles.  
For a bosonic field, the limit of high multiplicity corresponds to new physics that is displaced in  {\it field space} relative to the vacuum in which experimentalists and their experiments reside.   One can then subtly modify the laws of nature, for instance by introducing boundaries for the electromagnetic fields or the Higgs field, which are experimentally unconstrained simply because humans have not created or observed sufficiently  {\it large and coherent} fields.
More generally, we will argue that field space endpoints are a novel Lorentz invariant, local, and unitary deformation of any effective field theory (EFT).

The notion of new physics lurking at high multiplicity may seem contrived.  After all, shoehorning some ludicrously high dimension operator of the form $\phi^{n}$ for $n \gg1$ serves little purpose other than to generate, via quantum corrections, a tower of lower dimension operators that are more directly accessible  by experiment.

Fortunately, a more sensible incarnation of this phenomenon exists and is in fact commonplace in EFT.  For an example, consider a toy universe composed of a periodic scalar field $\phi$ with the oscillatory potential,
\eq{
V_{\rm osc}(\phi)  &= -\epsilon  \Lambda^4 \cos\left(\frac{\phi}{\Lambda}\right),
}{}
where $\epsilon \ll 1$.  An experimentalist in this world painstakingly constructs successively larger colliders, diligently extracting the coefficients of  higher dimension operators, $\epsilon \phi^6/\Lambda^2$, $\epsilon \phi^8/\Lambda^4$,
%, $\epsilon \phi^{10}/\Lambda^6$, 
etc., each pointing to a scale of new physics, $\epsilon^{-1/2} \Lambda $, $ \epsilon^{-1/4} \Lambda$, 
%$\Lambda \epsilon^{-1/6}$, 
etc.  These {\it mirage cutoffs} are parametrically higher than the true ultraviolet (UV) cutoff, which up to log corrections in $\epsilon$, is of order $\Lambda$ and only emerges at high multiplicity \footnote{This statement applies to the potential, but in general one must also grapple with low multiplicity derivative interactions that are set by the scale $\Lambda$ without any additional suppression by $\epsilon$.  In this case the true cutoff $\Lambda$ is accessible at low multiplicity and there is no mirage.  For example, unsuppressed interactions of this kind appear in many UV completions of the axion after integrating out an accompanying radial Higgs mode or tower of extra-dimensional excitations.   
Still, it is still striking that the axion potential exhibits a mirage cutoff.}.   In the context of periodic scalars, the idea of burying new physics at high multiplicity was studied in recent interesting work \cite{Hook:2023pba}.

An even more pedestrian incarnation of this phenomenon occurs in a finite, effectively one-dimensional solid---also known as a bar.  As reviewed in  \App{app:bar}, the phonon field $\phi$ is {\it bounded in field space} because the bar itself is finite.  Mechanically, this can be implemented by an infinite potential wall  \footnote{Alternatively, we can enforce a boundary on field space by multiplying the kinetic energy term by a step function $\theta(\bar \phi -\phi)$.  The physical equivalence between an infinite wall in the potential energy versus a step function in the kinetic energy is analogous to what happens in Einstein versus Jordan frame in gravity.
For a detailed review of the EFT of phonons in a physical bar, see \App{app:bar}.  In \App{app:balls_springs_strings} we also present a simple model of balls, springs, and strings whose mean field approximation exhibits a boundary in field space. },
\eq{
V_{\rm wall}(\phi)  &= \left\{ 
\begin{array}{ll}
0 &,\quad  |\phi| < \bar\phi\\
\infty &,\quad  |\phi| > \bar\phi
\end{array}
\right. ,
}{V_wall}
which enforces $|\phi| < \bar{\phi}$ for some positive $\bar \phi$.
While field space endpoints may seem exotic, we emphasize that they are next of kin to periodic boundary conditions.

Let us regulate $V_{\rm wall}(\phi) $ by introducing a finite wall thickness $\Lambda$ and considering an exponential potential,
%\footnote{This exponential potential only weakly stabilizes the vacuum to the origin.  
%%exhibits a runaway for the field. While this technically makes the vacuum unstable, it is easy to cancel this runaway by adding a small tadpole to the potential.   
%Obviously, since we are concerned about the physics of the wall and not the runaway, we will not concern ourselves here with the details of stabilizing the vacuum and simply assume that it is at $\phi =0$. },
\eq{
V_{\rm exp}(\phi)  &= \Lambda^4 \exp\left(\frac{\phi^2- \bar\phi^2}{\Lambda^2}\right).
}{V_exp}
As a series expansion, the potential is
\eq{
V_{\rm exp}(\phi) = \epsilon \Lambda^4  \sum_{n=0}^\infty \frac{1}{n!} \left(\frac{\phi^2}{\Lambda^2} \right)^n ,%=   \sum_{n=0}^\infty \frac{1}{n!} \frac{\phi^n}{\Lambda_n^{n-4}} ,
}{}
where $\epsilon = \exp(-\bar{\phi}^2/\Lambda^2)\ll 1$.   
%$\epsilon = e^{-\bar{\phi}^2/\Lambda^2}\ll 1$.  
Since $V_{\rm exp}(\phi)$ is basically a Wick rotation of $V_{\rm osc}(\phi)$, it also exhibits a mirage cutoff.

% Notably, the case of the infinitely steep wall $V_{\rm wall}(\phi)$ corresponds to $V_{\rm exp}(\phi)$ in the limit of {\it vanishing} cutoff $\Lambda \rightarrow 0$, but we will see that this is not pathological because this breakdown is entirely invisible to any process at finite multiplicity in the $\phi = 0$ vacuum.  Indeed, for $\Lambda \rightarrow 0$ the wall has no effect except on coherent field configurations near $\phi \sim \bar \phi$. %, which will be intrinsically strongly coupled.

Are the boundaries of field space robust under quantum corrections?    For the periodic scalar the answer is yes, since $V_{\rm osc}(\phi)$ is invariant under a discrete shift symmetry that  constrains its radiative corrections.  On the other hand, for the bounded scalar, $V_{\rm wall}(\phi)$ must also be radiatively stable, purely on physical grounds.  After all, solids do not spontaneously implode or explode in the presence of quantum effects.   This leads to a natural question: in the language of EFT, how precisely does the length of a bar remain stable under quantum corrections, and can such a mechanism be generalized \footnote{Notably, boundaries in {\it coordinate space} have long appeared in the context of gravity as geometric cutoffs or ``Dirichlet walls''.   Since these boundaries are not dynamical branes, they need not be  stabilized, but at the same time their consistency is not guaranteed.  Nevertheless, under certain conditions, these endpoint objects have been found to be stable \cite{Andrade:2015gja,Andrade:2015qea,Marolf:2012dr} and consistent with the initial value problem \cite{An:2021fcq}. }?

We will argue here that periodic and bounded scalars are but a subset of a much broader class of EFTs that exhibit field space boundaries and mirage cutoffs.  These EFTs are defined by a potential of the general form \footnote{Here we define our potential to be a series expansion in $(\phi^2/\Lambda^2)^n$ with coefficients that scale as $1/n!$, as opposed to a series expansion in $(\phi/\Lambda)^n$ with coefficients that scale as $1/n!$.  We do this for two reasons.  Firstly, our phenomenological applications all involve field space boundaries for charged fields like the Higgs boson or the SM force carriers, which only appear in even powers in their self-interactions.  The second reason has to do with a minor subtlety regarding the scale of unitarity violation.  A general potential defined by a series expansion in $(\phi/\Lambda)^n$ with coefficients $1/n!$ will cover the familiar case of a periodic scalar, as well as the exponential runaway of a dilaton field.   However, as shown in \cite{Hook:2023pba}, these models exhibit scattering which at {\it any} multiplicity will never see a breakdown of the EFT at the scale $\Lambda$ but rather at the slightly higher cutoff $\Lambda \sqrt{\log(1/\epsilon)}$, which can be realized in explicit UV models \cite{Hook:2023pba}.  Of course, these factors of $\log(1/\epsilon)$ corrections are far less important than the powers of $\epsilon$ that generate the phenomenon of the mirage cutoff.  Nevertheless, for simplicity we restrict here to a potential which is a series expansion in $(\phi^2/\Lambda^2)^n$ with coefficients $1/n!$, which as shown in \App{app:scattering} has a physical cutoff defined by unitarity violation in scattering that is precisely the scale $\Lambda$, with no additional factors of $\log(1/\epsilon)$.   Note that in general, changing the power of $\phi$ in the series expansion has the same effect as changing the scaling of the series coefficient.  So for example, a series expansion in $(\phi^k/\Lambda^k)^n$ with coefficients $1/n!$ is the same as a series expansion in $(\phi/\Lambda)^n$ with coefficients $(1/n!)^{1/k}$. }, 
\eq{
V(\phi) %&= \epsilon \times (\textrm{entire function of } \phi/\Lambda) \\ 
&= \epsilon \Lambda^4  \sum_{n=0}^\infty \frac{c_n}{n!} \left(\frac{\phi^2}{\Lambda^2} \right)^n ,
}{V_gen}
where $c_n \sim {\cal O}(1)$ are {\it arbitrary}.  This potential---which is just an entire function of the field times a small spurion---is radiatively stable and induces a boundary at
\eq{
\bar \phi \sim \Lambda \sqrt{\log (1/\epsilon)}.
%\epsilon = e^{-k(\bar\phi/\Lambda)^{1/k} } .
}{barphi}
%This potential is generated by UV modes at the true cutoff $\Lambda$, but with additional suppression by a spurion $\epsilon\ll1$.  
We will also show that $2n$-point scattering violates unitarity when the energy per particle exceeds the multiplicity-dependent mirage cutoff,
\eq{
\Lambda_n \sim \epsilon^{-1/2n } \Lambda ,
}{Lambda_n}
which at finite $n$ is much larger than the true cutoff $\Lambda$.  The mirage dispels for $n \gg n_*$, where we have defined
%To obtain the true cutoff of the EFT, we minimize $\Lambda_n$ as a function of $n$, yielding the critical multiplicity%is the effective {\it multiplicity-dependent} cutoff seen by each operator at $n>4$. 
\eq{
n_* \sim \tfrac12 \log(1/\epsilon) \sim \tfrac12 \bar\phi^2/\Lambda^2 \gg 1,
%  \log(1/\epsilon)= \bar \phi/\Lambda,
%\lim_{n\rightarrow\infty} \Lambda_n = \Lambda,
}{n_crit}
at which point the true cutoff of the EFT emerges.
%\eq{
%\Lambda_* = \textrm{min}( \Lambda_n) \sim \log^{k-1/2}(1/\epsilon) \Lambda.
%}{Lambda_true}
%The hierarchy $\Lambda_*/ \Lambda_n \ll1$ precisely quantifies the phenomenon of the mirage cutoff.  
Since field space boundaries are only visible at high multiplicity, they are very weakly constrained observationally, allowing for boundaries of the Higgs and force carrier fields of the SM associated with exceedingly low cutoffs.

%In \Sec{sec:boundary} we show how $V(\phi)$ induces a field space boundary and determine the regime of validity of the resulting theory in \Sec{sec:regime}.  Afterwards, we show in \Sec{sec:naturalness}  that $V(\phi)$ is radiatively stable, in the sense that quantum corrections can modify the numerical coefficients of $V(\phi)$ but not its general structure.  In   \Sec{sec:BSM} we then describe the application of bounded field space to the Higgs field and the long-range force carriers of the SM.   Lastly, in \Sec{sec:discussion} we discuss future directions.

\medskip

\section{Field Space Boundary}
\label{sec:boundary}

Even though the potential $V(\phi)$ in \Eq{V_gen} is very general, it exhibits a universal behavior: a long, flat plateau interrupted by a steep rise or fall in the potential near $|\phi|\sim \bar\phi$.  This pattern is easily checked by plotting $V(\phi)$ for random values of $c_n\sim {\cal O}(1)$.  
Intuitively, this structure forms because at small field values, $|\phi| \ll \bar\phi$, the potential is dominated by $\epsilon$ and thus nearly vanishing.  Near $|\phi| \sim \bar{\phi}$, the field finally overpowers $\epsilon$, yielding swift growth.
Note that the equation for $\bar\phi$ in terms of $\epsilon$ in \Eq{barphi} is derived by identifying the largest term in \Eq{V_gen} as a function of $n$ at fixed $\phi$ and then solving for the value of $\phi$ for which that term is of order $ \Lambda^4$.

%\eq{
%\bar \phi = \Lambda \log(1/\epsilon).
%}{barphi}

There is a more precise argument for this universality.   Given that $c_n\sim {\cal O}(1)$, the Cauchy-Hadamard formula implies that the radius of convergence of $V(\phi)$ is infinite,
\eq{
\lim_{n\rightarrow\infty} \frac{\log |c_n/n!| }{ n} \sim - \lim_{n\rightarrow\infty}  \log n = -\infty,
}{}
so  $V(\phi)$ is entire \footnote{It may be concerning that this potential---like that of the axion---is an entire function.  In particular, there is evidence that the generating functional of connected correlators $W[J]$ cannot be entire \cite{Panagopoulos:2020sxp,Cohen:2022clv}, which is related to the asymptotic nature of the perturbative series \cite{Brown:1992ay,Son:1995wz,Rubakov:1995hq,Libanov:1997nt}.  Furthermore, in certain explicit examples the operator product expansion can also be asymptotic \cite{Dunne:1999uy}. Nevertheless, we note that these degenerations at high multiplicity all stem from the growth of diagrammatic topologies in {\it correlators}, while the potential defines individual {\it interaction vertices}.  As shown in \App{app:scattering}, any contribution to a correlators involving multiple insertions of the potential interactions will be subleading since  $\epsilon\ll 1$.  Consequently, the growth in diagrams is tamed and any subtleties concerning the asymptotic nature of perturbation theory are not relevant. }.    By Picard's little theorem, $V(\phi)$ must take on all possible values in the complex plane, modulo perhaps a single point.  Thus $V(\phi)$ must blow up for some complex value of $\phi$.  Note that for our purposes, this blow up need not exhibit a formally infinite singularity, but merely that $V(\phi)$ becomes much larger than $\Lambda^4$.
%Where does this blow up occur?  If $\epsilon=1$, then $\Lambda$ is the only parameter in the potential and the blow up happens near $\phi \sim \Lambda$.  However, our setup requires $\epsilon \ll 1$, so the blow up is postponed to $\phi \sim \bar \phi \gg \Lambda$.
For a periodic scalar, this happens at imaginary values of the field, while  for a generic $V(\phi)$ it is typically on the real axis.   If $V(\phi)$ is monotonically growing, for example if $c_n >0$, then this steep rise produces a potential wall that induces a field space boundary.  
Of course, there exist many potentials, for example with coefficients $c_n$ of indefinite sign, that oscillate wildly between walls and cliffs at $|\phi| \sim  \bar{\phi}$.  Since these potentials are oscillatory we will not consider them further \footnote{As shown in \Sec{sec:naturalness}, radiative corrections serve only to renormalize the coefficients $c_n$ to numerical values of the same order.  Since this manifestly preserves the structure of the potential in \Eq{V_gen}, it also maintains the flatness of the potential for $|\phi| \lesssim \bar \phi$.  Nevertheless, these corrections could in principle still flip the signs of $c_n$ in such a way that destabilizes the potential, toggling the feature at $|\phi| \sim \bar\phi$ from a wall into a cliff.  This is reminiscent of the sensitivity of vacuum alignment to quantum corrections, for example as occurs in radiative electroweak symmetry breaking.  Here also, the binary question of whether the potential develops a wall or a cliff will be decided by the UV dynamics that regulate the divergences, so in this sense these two possibilities have even odds.  Notably, if the underlying theory is supersymmetric, then the potential energy will be automatically bounded from below.}.

We emphasize that the above arguments do not rely on precise values for $c_n$, so field space boundaries are generic and robust under small corrections.

\section{Regime of Validity}
\label{sec:regime}

What is the regime of validity of the EFT defined by $V(\phi)$?  
Clearly, this theory breaks down for coherent field configurations near the boundary, $|\phi| \sim \bar \phi$ \footnote{One should bear in mind that the EFT transitions from the weak field to strong field
 regime long before $\phi$ reaches $\bar\phi$.  To see why, recall that the weak field regime by definition applies when perturbation theory in the field value is accurate.  This implies that the $n$-point interaction in $V(\phi)$ must be more important than the $(n+1)$-point interaction.  From  $V(\phi)$ it is clear that this transition occurs at the critical field value $|\phi| \sim \Lambda \ll \bar\phi$.}.  On the other hand, if the background value of $\phi$ is far from the boundary, for example at $\phi=0$, then the failure of the EFT is far more subtle because its only imprint is in the high multiplicity interactions of the fluctuations about this background. 
At low multiplicity, the center-of-mass energy $E$ and energy per particle $E/2n$ are relatively similar, but at high multiplicity they are completely different.  Hence, the cutoff of the EFT is not just specified by a characteristic energy---we must very precisely distinguish the total energy versus energy per particle.

%, whose effective cutoff is $\Lambda$ on account of \Eq{Lambda_n}.  
As shown in \App{app:scattering}, unitarity is violated when the energy per particle exceeds $E/2n \gg \Lambda_n$, where the mirage cutoff $\Lambda_n$ is defined in \Eq{Lambda_n}.   For low multiplicities, $n \ll n_*$, the mirage cutoff is parametrically higher than the true cutoff, so $\Lambda_n \gg \Lambda$.  At high multiplicities, $n\gg n_*$, we see that $\Lambda_n \sim \Lambda$ and the physical cutoff of the EFT emerges. This makes sense because the $2n$-point interaction in the potential is only activated when all $2n$ particles in the process are imbued with sufficient energy and multiplicity to probe the boundary. Since the energy of each particle cannot exceed $\Lambda$ while remaining within the EFT, we conclude that this scale is the true cutoff. All divergent loop integrals should be cut off by $\Lambda$ and moreover new physics must appear at $\Lambda$ to restore unitarity, which will be very important for radiative stability.  Lastly, since the true cutoff only emerges at $n_* \sim \tfrac12 \log(1/\epsilon)\sim \tfrac12 \bar \phi^2/\Lambda^2$, unitarity violation occurs when the center-of-mass energy exceeds $E \gg  \bar \phi^2 /\Lambda$.

%What is the precise multiplicity at which the EFT breaks down?  
Let us examine the break down of the EFT in some examples. %Here the answer depends on the microphysics.  In particular, in our discussion of the potential $V(\phi)$, we introduced both  the spurion $\epsilon$ as well as the field space boundary $\bar \phi$.  These are interchangeable parameters, related to each other by \Eq{barphi}.  Depending on the underlying model, one should either fix $\epsilon$ or $\bar \phi$.  
For periodic scalars it is quite common to encounter a nonperturbatively generated potential, in which case the spurion $\epsilon \sim e^{-4\pi/\alpha} \ll 1$ is exponentially small.
%sensitive to the gauge coupling in the UV and essentially independent of $\Lambda$ modulo threshold effects.  In this case, extraordinarily small couplings are relatively generic: for 
For example, for $\alpha \sim 1/137$ set by the fine structure constant, the true cutoff emerges at enormously high multiplicities defined by $n\gg n_* \sim  861$.
% Here the mirage cutoff is as defined in \Eq{Lambda_n} and the true cutoff emerges for multiplicities satisfying \Eq{n_crit}.    

 Meanwhile, for a physical bar we have that $\epsilon \sim \exp(-\bar{\phi}^2/\Lambda^2)$ for fixed $\bar \phi$.    %Approximating the end of the bar with the exponential potential in \Eq{V_exp}, we find that the cutoff $\Lambda$ controls the gradient of the wall, while $\Lambda_n = \Lambda e^{\bar\phi/\Lambda(n-4)}$ is the mirage cutoff.  
 The limit of an infinitely steep transition corresponds to a vanishing cutoff $\Lambda \rightarrow  0$.  While this is naively pathological, at fixed multiplicity this limit also sends the mirage cutoff to $\Lambda_n \rightarrow \infty$, so by \Eq{n_crit} the boundary can only be detected by processes that are of formally infinite multiplicity. 
This entirely decouples the boundary from observables in the trivial vacuum, so it can only be accessed by coherent fields.

%Said another way, even though the cutoff $\Lambda \rightarrow  0$ is vanishing, 
%\Eq{n_crit} implies that 
%  Instead keeping $\Lambda$ small but finite, this true cutoff only emerges in processes at or beyond the critical multiplicity defined in \Eq{n_crit}.

\section{Radiative Stability}

\label{sec:naturalness}

The form of $V(\phi)$ is stable under quantum corrections, which in certain cases follows from symmetry.  The periodic potential, $V_{\rm osc}(\phi)$, is invariant under  $\phi \rightarrow \phi + 2\pi k \Lambda$ for integer $k$, while the wall potential $V_{\rm wall}(\phi)$ is invariant under $\phi \rightarrow \phi + \mu \theta(\bar\phi - |\phi|)$ for infinitesimal $\mu$.

Remarkably, any potential of the form of $V(\phi)$ is radiatively stable in a broadened sense: loops will correct the potential in a way that renormalizes its numerical coefficients by ${\cal O}(1)$ factors but otherwise preserves its general structure.  %To see why,
%coefficients $c_n$ but the parametric size of each term in the potential will be unchanged.  Since the emergence of a field space boundary does not require precise values for $c_n$, we see that the boundary is radiatively stable, as expected.
%To see why, let us
%consider the divergent loop corrections involving the potential.    
As shown in \App{app:scattering}, a quadratic divergence generated by a loop of $\phi$ is regulated by $\Lambda$ since that is the scale at which new physics emerges to restore unitarity.  Hence, the one-loop diagram involving the $2n$-point vertex $\epsilon \Lambda^4 (\phi^2/\Lambda^2)^n$ generates a contribution to the $2(n-1)$-point vertex of the form $\epsilon \Lambda^4 (\phi^2/\Lambda^2)^{n-2}$.  Since a term of this form is already accounted for in $V(\phi)$, this one-loop correction simply  renormalizes the corresponding numerical coefficient.  The same logic then applies to the $2(n-2)$-point vertex and so on and so forth.  %Hence, perturbative quantum corrections leave the form of $V(\phi)$ unchanged since $\Lambda$ is the only scale in play.  Here we have been sloppy with $n$-dependent numerical factors, but 

Of course, the above argument is glib about $1/16\pi^2$ loop suppression and $n$-dependent enhancement from Wick contractions.
Instead of estimating every conceivable diagram, let us compute all radiative corrections in one go using the Coleman-Weinberg (CW) potential, 
\eq{
V_{\rm CW}(\phi) =\frac{\Lambda^2 M^2(\phi) }{32\pi^2} + \frac{M^4(\phi)}{64\pi^2}  \log\left( \frac{M^2(\phi)}{\Lambda^2} \right) +\cdots,
}{}
where $M^2(\phi) =  V''(\phi)$
% $M^2(\phi) = U''(\phi) = V''(\phi) + \tilde V''(\phi)$ 
is the field-dependent mass for the hard scalar quanta
and the ellipses denote contributions beyond one loop.
As before, all divergences are regulated by $\Lambda$, which is the true cutoff at which new physics intervenes.  The all-loop CW potential depends on the field-dependent interaction vertices of the hard quanta, which at two-, three-, four-point are $V''(\phi)$, $V'''(\phi)$, $V''''(\phi)$, and so on. Since these interactions are vanishingly small for $|\phi| \ll \bar \phi$, the all-loop CW potential does not lift the flat region of the potential, confirming that field space boundaries are radiatively stable.  
%Note that since $V(\phi)$ and its derivatives are entire functions, so too is the CW potential, modulo log terms reflecting the choice of renormalization scale.  

%Furthermore, we show in \App{app:power} that for general $V(\phi)$, the all loop effect of including power law divergent corrections is to 
%apply a Gaussian {\it diffusion filter} to $V(\phi)$ which blurs its features in field space at scales shorter than $\Lambda/4\pi$.    preserving the structure of $V(\phi)$ as well as its field space boundaries.

Since $\epsilon \ll 1$, the leading radiative corrections all involve diagrams with a single insertion of the potential.   
Working at linear order in $\epsilon$, the renormalized potential coming from power law divergences {\it at all loops} is
\eq{
V_{\rm div}(\phi) &= \exp\left(\frac{\Lambda^2}{32\pi^2} \frac{d^2}{d\phi^2}\right) V(\phi)\\
&=\frac{\sqrt{8\pi} }{\Lambda}\int_{-\infty}^\infty d\phi'   \exp\left(-\frac{8\pi^2(\phi'-\phi)^2}{\Lambda^2}\right) V(\phi'),
}{V_div}
where $d^2/d\phi^2$ generates quadratically divergent loops that each evaluate to $\Lambda^2/16\pi^2$.    \Eq{V_div} implies that the renormalized potential $V_{\rm div}(\phi) $ is just the bare potential $V(\phi)$ convoluted % over a Gaussian kernel whose variance is $\Lambda/4\pi$.  Hence, these corrections act as a 
with a Gaussian {\it diffusion filter} that blurs any features in field space shorter than $\Lambda/4\pi$.    Since the potential wall has thickness $\Lambda$, this preserves the location of the field space boundary.  Note also that the Gaussian kernel will always map the entire function $V(\phi)$ to another entire function $V_{\rm div}(\phi)$.  In \App{app:power} we show in an explicit example how all loop power divergences preserve the size of the series coefficients, so $c_n \sim {\cal O}(1)$.

% loops. This establishes that the radiative corrections preserve the size of the numerical coefficients in the potential to be $c_n \sim {\cal O}(1)$.

%Now the key point is that $V(\phi)$ only enters into this formula through its second derivative, $V''(\phi)$.  Moreover, in the flat region $\phi \ll \bar\phi$, this second derivative is essentially zero, $V''(\phi)$.   Thus the one-loop CW potential does not produce corrections in the region  $\phi \ll \bar\phi$.

Bear in mind that the potential interactions in this EFT will generate derivative interactions  of the form \footnote{The potential terms generate derivative interactions in the same way as in $\phi^4$ theory, where wavefunction renormalization appears at two loops. }
\eq{
 \epsilon  \Lambda^4  \sum_{n=0}^\infty \sum_{m=0}^\infty \frac{c_{n,m}}{n! m!} \left(\frac{\phi^2}{\Lambda^2}\right)^n \left(\frac{\partial\phi\partial\phi}{\Lambda^4}\right)^m .
}{}
As usual, we should expect all operators consistent with symmetries---including derivatives---with all dimensionful quantities written in units of the cutoff $\Lambda$ and suppressed by $\epsilon$ \footnote{What is the natural size for these derivatively coupled interactions?  Consider, for example,  the $ (\partial \phi)^4$ operator, whose coefficient is famously constrained to be nonnegative by analyticity, unitarity, and causality \cite{Adams:2006sv}.  In particular, this coefficient is equal to an energy weighted integral over the total cross-section for $\phi \phi \rightarrow \chi$, where $\chi$ denotes the heavy states in the UV completion.  To preserve the phenomenon of the mirage cutoff, the coefficient of $ (\partial \phi)^4$  must be suppressed by $ \epsilon \ll 1$, which implies that $\chi$  must couple very weakly to a pair of $\phi$ particles.  This is very different than what happens in traditional UV completions, like the linear sigma model completion of a periodic scalar.   Conversely, an EFT with a mirage cutoff can only arise in a UV completion in which the heavy state $\chi$ only couples to a {\it high multiplicity} of $\phi$ particles.   }. Naively, these derivative interactions induce boundaries not just for $\phi$, but also for $\partial \phi$, $\partial^2\phi$, etc.  
However, since the energy per particle  is less than $\Lambda$ within the EFT, the latter boundaries are not physical.

\section{Hierarchies}

\label{sec:hierarchy}

Up until now we have considered the very simplest case in which the whole potential is $V(\phi)$.  
However, the plot thickens if we consider the extended potential,
%\eq{
$U(\phi) = V(\phi)+ \tilde V(\phi)$,
%}{}
where $\tilde V(\phi)$ is a set of arbitrary, possibly higher dimension operators.    Importantly, the contribution from $V(\phi)$ generates a wall at $|\phi| \sim \bar\phi$, so this extended potential still exhibits a field space boundary.    Furthermore, while $\tilde V(\phi)$ may have sizable interactions, all contributions to the CW potential involving $V(\phi)$ or its derivatives will be suppressed for $|\phi| \ll \bar \phi$, so like before the field space boundary is stable.

In general, there is another cutoff $\tilde \Lambda$ associated with the higher dimension operators and UV divergences in $\tilde V(\phi)$.   Physically,  $\tilde\Lambda$ is a proxy for the scale of new physics that UV completes $\tilde V(\phi)$.   Since $\phi$ is a scalar, it suffers from a ``hierarchy problem'' \footnote{If the mass of $\phi$ is calculable as a function of the parameters of the UV theory, then the dependence on $\tilde \Lambda$ will emerge from the physical mass scale of the heavy degrees of freedom.  This happens, for example, in phenomenological models of soft supersymmetry breaking.  If the mass of $\phi$ is incalculable, however, then  its quadratic divergences must cancelled by a counterterm whose value is uniquely fixed by the physical mass, which can only be extracted from experimental observations.  In this case, the mass cannot be predicted in terms of UV parameters, and technically there is no hierarchy problem.} because its mass receives quadratically divergent corrections of the form $\tilde\Lambda^2/ 16\pi^2$.  

If the resulting hierarchy problem is independent of the microphysics associated with the field space boundary, then $\tilde\Lambda$ and $\Lambda$ are independent quantities.   
  Furthermore, if $\tilde\Lambda\ll \Lambda$, then the true cutoff $\Lambda$ of the boundary potential---as well as all of the associated mirage cutoffs---is parametrically higher than $\tilde\Lambda$.   Even if the boundary potential $V(\phi)$ receives loop corrections cut off by $\tilde \Lambda$, these contributions are subleading to those cut off by $\Lambda$, so the field space endpoint is robust.
  %This leads to the uninteresting case in which the physics generating the boundary potential is at a higher energy scale than the physics that resolves the hierarchy. 
  On the other hand, if $\tilde \Lambda\gg \Lambda$ then one might worry that loops that are quadratically divergent in $\tilde \Lambda$ could destabilize the boundary potential $V(\phi)$.  This is avoided, however, since whatever mechanism cancels the quadratically divergent contributions to $\phi$ mass---be it supersymmetry, compositeness, or simply fine-tuning---also cancels the corresponding divergences in $V(\phi)$, diagram by diagram.

A very speculative but also interesting possibility is that the boundary is linked to the hierarchy problem, in which case $\tilde \Lambda\sim \Lambda$.  We return to this possibility later on.

\section{Beyond the Standard Model}

\label{sec:BSM}

%Let us now consider how the SM can be modified to include field space boundaries.

\smallskip
\noindent{\it Higgs Field.}  The Higgs field  $H$ plays a central role in the SM.   Its vacuum expectation value (VEV) has been inferred to be $\langle H\rangle  \sim 246$ GeV from nearly a century of experiments probing the masses and interactions of the quarks, leptons, photons, and electroweak gauge bosons \cite{ParticleDataGroup:2024cfk}.  The Higgs VEV is far and away the largest field configuration ever to be accessed experimentally by humans, albeit indirectly.  The fluctuations of the Higgs field about its VEV are the Higgs particle, whose existence was spectacularly confirmed at the LHC \cite{ATLAS:2012yve,CMS:2012qbp}.

In spite of this, current experiments have not accessed coherent Higgs fluctuations about its VEV.  The LHC has only sensitively probed interactions involving a single Higgs boson.  More generally, low multiplicity interactions will at best reveal the local curvature of Higgs field space \cite{Alonso:2015fsp}.     As long as the boundary in Higgs field space is larger than $\langle H\rangle$, it will be experimentally invisible, appearing only in high multiplicity Higgs interactions.  %Note that the local curvature of Higgs field space near the origin can be inferred from measurements at low multiplicity \cite{Alonso:2015fsp}.

This all offers motivation to modify the field space of the Higgs and ask about its possible experimental repercussions.   Our EFT analysis has shown that a gauge invariant potential of the general form
\eq{
V (H) &= \epsilon  \Lambda^4  \sum_{n=0}^\infty \frac{c_n}{n!} \left(\frac{|H|^2}{\Lambda^2}\right)^n ,
}{}
induces a field space boundary and is radiatively stable.
%Here we recall that the interactions in $V(H)$ induce a breakdown of the EFT in high multiplicity processes at energies $E\gtrsim \Lambda$.  This implies that $\Lambda$ is the approximate scale at which new physics must appear.
%Assuming that divergent loop integrals involving $V(H)$ are regulated by $\Lambda$, we see again that this potential is radiatively stable.
Concretely, consider $U(H) = V(H)+ \tilde V(H)$, where
  \eq{
  V(H) &=\Lambda^4 \exp\left(\frac{|H|^2- \bar H^2}{\Lambda^2}\right)\\
  \tilde V(H) &=\frac{\lambda}{4} |H|^4 -m^2 |H|^2,
  }{}
  has an exponential wall with $\epsilon = \exp(-\bar H^2/\Lambda^2)\gg 1$, together with the usual SM potential.
Electroweak symmetry breaking that makes direct use of a Higgs boundary does not appear to be viable \footnote{
Is there a version of electroweak symmetry breaking that directly utilizes a boundary in Higgs field space?  The simplest version of this idea is to set $\lambda=0$, so the tachyonic mass for the Higgs field drives its VEV to the boundary of field space. 
 For $\Lambda \ll \bar H$, which is required for radiative stability of the boundary,  the VEV is pushed to the edge, $\langle H\rangle \sim \bar H$, as expected, irrespective of the value of tachyonic mass $m^2$.  Unfortunately, by expanding in fluctuations, $H = \langle H\rangle + h$, we see that the Higgs particle exhibits nonlinear interactions that enter in powers of $h \bar H/ \Lambda^2$, which are enhanced for $\Lambda \ll \bar H$.  Since the  Higgs VEV is pinned inside the cusp between the tachyonic mass and the boundary, the Higgs particle is strongly coupled and thus not viable.}, so we will assume hereafter that $\bar H > \langle H\rangle $ is safely above the VEV.  Of course,  such a boundary could still play a role in the strength of the electroweak phase transition, which is critical for models of baryogenesis.

Let us comment on possible connections between field space boundaries and the electroweak hierarchy problem.  The Higgs mass parameter receives quadratically divergent corrections, $m^2 \sim \tilde\Lambda^2/ 16 \pi^2$, indicating sensitivity to physical states at the UV cutoff $\tilde\Lambda$ associated with the SM.  We  saw earlier what happens when $\tilde\Lambda$ and $\Lambda$ are independent scales.  The hierarchy problem is independent of the boundary dynamics and consequently the field space endpoints are radiatively stable.

A more exotic option is that $\tilde \Lambda \sim \Lambda$ are linked.  
 This offers a solution to the hierarchy problem reminiscent of large extra dimensions \cite{Arkani-Hamed:1998jmv}: if the true cutoff of the SM is $\Lambda \sim 1$ TeV then the Higgs mass parameter $m^2$ is actually {\it natural}.    However, unlike the scenario of large extra dimensions, the degrees of freedom associated with a boundary potential would be exceedingly difficult to observe. In particular, they would have evaded detection simply because they only appear in very high multiplicity interactions that are inaccessible to direct collider probes and indirect signatures from precision electroweak observables or flavor violation.  By the same token, new physics of this kind should couple exceedingly weakly to the Higgs bilinear $|H|^2$, so it strains the imagination to conceive of how those degrees of freedom could ever cancel the divergences of the Higgs at a mechanical level.  Hence, such a scenario seems implausible.

\smallskip

\noindent{\it Force Fields.} %An especially drastic modification of the SM is to introduce field space boundaries for the force carriers: the photon, gluon, and graviton.  
Next, consider the case of a boundary for the electromagnetic field.   A version of this idea has existed for almost a century in the form of the Born-Infeld (BI) model, which is a nonlinear generalization of electromagnetism with the  Lagrangian,
\eq{
L(F) &=- \Lambda^4 \sqrt{-\det(\eta_{\mu\nu} +F_{\mu\nu}/\Lambda^2)	}.
}{}
As is well-known, BI theory regulates the Coulomb divergence of point charges by imposing a fundamental limit on the electromagnetic field strength.  In BI theory, the parameter  $\Lambda$ plays dual roles as the boundary of field space {\it and} the energy scale at which photon scattering at any multiplicity violates unitarity.

Instead, let us consider the general Lagrangian for the electromagnetic field strength \footnote{We have imposed a boundary on the electromagnetic field so as to preserve gauge invariance.   In principle, we could instead consider an infinite wall boundary on the gauge field itself, corresponding to a Proca potential which has been sequestered to high multiplicity.  If we regulate the boundary wall to have finite thickness, then tiny but nonzero Proca potential terms will be induced in the trivial vacuum.   While such a setup might be viable for an abelian gauge field like hypercharge, this is likely pathological for the nonabelian case, where Proca mass terms invariably lower the cutoff. },
\eq{
L (F) &= \epsilon  \Lambda^4  \sum_{n=0}^\infty \frac{c_n}{n!} \left(\frac{F^2}{\Lambda^4}\right)^n.
% \, {\rm tr} \left(\frac{F_{\mu\nu}}{\Lambda^2}\right)^n,
}{LF}
%Here a natural gauge invariant object to bound is $F_{\mu\nu} F^{\mu\nu}$,
% = -2( \vec{E}^2- \vec{B}^2), 
%but unfortunately this quantity is not positive definite.  Instead we consider the potential
%\eq{
%V (F) &= \epsilon  \Lambda^4  \sum_{n} \frac{c_n}{n!} \left(\frac{(F_{\mu\nu} F^{\mu\nu} )^2}{\Lambda^8}\right)^n,
%}{VF}
%which induces a field space boundary on the positive definite quantity $(F_{\mu\nu} F^{\mu\nu})^2$.  In analogy with \Eq{barphi}, this quantity is bounded by $\bar F^4= \Lambda^8 \log(1/\epsilon)$.
%\Eq{VF} is naively reminiscent of 
In the SM, this Lagrangian should technically be defined in terms of the hypercharge or electroweak gauge bosons, but we can ignore this caveat for now.
 As shown in \App{app:LF}, %the small spurion is $\epsilon \sim e^{-\bar F^2/\Lambda^4}$, 
 the mirage cutoff is $\Lambda_n \sim \Lambda \epsilon^{-1/4n}$, the boundary for the electromagnetic field is induced at $\bar F= \Lambda^2 \sqrt{\log(1/\epsilon)}$, and the critical multiplicity at which this endpoint can be detected corresponds to
%  $n\gg \frac12 \log(1/\epsilon) = \frac12 \bar F/ \Lambda^2$.  
  $n\gg n_* \sim \frac14 \log(1/\epsilon) \sim  \tfrac14 \bar F^2/ \Lambda^4$ \footnote{Strictly speaking, the number of photons produced in a given scattering process is ill-defined on account of infrared divergences.  However, since the field strength associated with a soft photon is vanishingly small, it essentially decouples from all higher dimension operators.  This accords with the calculation in \App{app:LF}, which shows that EFT breaks down when the energy per particle exceeds the cutoff.}.  
   For processes at this multiplicity, unitarity is violated when the energy per particle exceeds $E/2n\gg \Lambda$, which corresponds to a center-of-mass energy $E \gg  \frac12 \Lambda \log(1/\epsilon) \sim \frac12 \bar F^2 / \Lambda^3$.
%    For completeness, in \App{app:LF} we present an explicit example of a Lagrangian of the form of \Eq{LF} that exhibits a boundary for the electromagnetic field.

What values of $\Lambda$ and $\bar F$ are phenomenologically permitted?   In principle, this setup is constrained cosmologically by  the successful predictions of big bang nucleosynthesis, which are certainly untouched if the cutoff is higher than the temperature, $\Lambda \gtrsim 1 \textrm{ MeV}$.  It is a detailed question whether the opposite regime is ruled out.  

 Meanwhile, 
a leading constraint from collider physics is from light-by-light scattering induced by the interaction $\epsilon F^4/ \Lambda^4$.    Measurements of relativistic heavy-ion collisions at the LHC imply a lower bound on the effective cutoff of the $F^4$ operator of order $\sim 100 \textrm{ GeV}$ \cite{Ellis:2017edi,ATLAS:2017fur,dEnterria:2013zqi}, which translates into the constraints,
\eq{
\Lambda \gtrsim \epsilon^{1/4}  (100 \textrm{ GeV}) \quad \textrm{and} \quad
\bar F \gtrsim \epsilon^{1/2}  \log(1/\epsilon) (100 \textrm{ GeV})^2,
}{LBL}
which are easily evaded by cranking down $\epsilon \ll 1$.

For example, consider the case of a preposterously low cutoff $\Lambda \sim 5\textrm{ GeV}$ with a tiny spurion $\epsilon \sim e^{-2000}$, which can easily arise nonperturbatively.   A cutoff at this scale is naively ruled out by experiment---after all, we have measured a litany of SM processes and particle masses that clearly exceed $\Lambda$. However, the crucial point is that $\Lambda$ only emerges in $2n$-point processes that exceed the critical multiplicity of $2n \gg   1000$ photons.  This implies that the EFT breaks down for center-of-mass energies exceeding $E \gg 5 \textrm{ TeV}$.    Processes below this threshold energy are simply kinematically forbidden from producing a sufficient multiplicity of photons that have sufficient energy to violate unitarity in the EFT.  

In this example the mirage cutoff associated with the $F^4$ operator is $\Lambda  \epsilon^{-1/4}  \sim 10^{218} \textrm{ GeV}$, which is obviously safe from the bounds in \Eq{LBL}.  We leave for future work how these models are constrained by limits on ``soft bombs'' at the LHC \cite{Knapen:2016hky,CMS:2024nca}.
 Furthermore, the bound on the electromagnetic field is $\bar F \sim (33 \textrm{ GeV})^2$.  This is still much higher than the strongest electromagnetic fields ever experimentally observed, like those in magnetars or planned beam experiments \cite{Abramowicz:2019gvx}. Note that a field of this size is screened by Schwinger pair production, so a stable electromagnetic field at the boundary can only exist in models where  $\bar F$ is below the $e^+e^-$ mass threshold.

% For $\Lambda$ and $\bar F$ that saturate the light-by-light scattering bounds in \Eq{LBL}, the Schwinger limit constraint in \Eq{Schwinger} implies that $\epsilon \gtrsim 2.5\times  10^{-26}$.  By construction the mirage cutoff is $\epsilon^{-1/4} \Lambda \sim 100 \textrm{ GeV}$, while the true cutoff $\Lambda \sim 130$ keV is extremely low but only emerges  at multiplicities of order $n \gg \bar F /\Lambda^2 \sim 58$ photons \footnote{Strictly speaking, the number of photons produced in a given scattering process is ill-defined on account of infrared divergences.  However, since the linearized field strength associated with a soft photon is vanishingly small, it essentially decouples from all higher dimension operators.  Conversely, unitarity violation in these operators can only probed by hard photons, so it is their multiplicity $n$ which is relevant for the true cutoff.  }. '

Similar logic applies to field space boundaries for the gluon field, which are implemented by a Lagrangian like \Eq{LF} but for the gluon field strength.  The leading constraint from the LHC applies to the anomalous gluon trilinear coupling $G^3$, while an unscathed QCD phase transition would require $\Lambda$ and $\bar G$ to be above the QCD scale.  Last but not least, a boundary in the context of gravity can be induced by a version of \Eq{LF} for the Riemann curvature.  Since direct constraints on $R^3$ are exceedingly weak, $\Lambda$ and $\bar R$ can again be very low.

\section{Discussion}
\label{sec:discussion}

A boundary in field space introduces new physics, possibly at a very low scale, that is sequestered from the vacuum and invisible except through coherent field configurations and high multiplicity interactions. 
These are very weakly constrained by experiments, so field space boundaries are a viable Lorentz invariant, local, and unitary deformation of the SM.  
 %Intuitively, boundaries introduce new physics at a very low scale that is sequestered in field space. 
 As one would expect from the very existence of finite solids, field space boundaries are radiatively stable and robust under small corrections.   This work leaves several avenues for future exploration.

First and foremost is the challenge of devising a fully UV complete theory which exhibits a field space boundary and its accompanying features at high multiplicity.  Seemingly related ideas have appeared in string theoretic models of inflation \cite{Panagopoulos:2019ail,Panagopoulos:2020sxp,DeLuca:2021pej} that generate observationally relevant tails in high multiplicity correlation functions.
It would be illuminating to explore whether these UV completions  preserve the phenomenon of the mirage cutoff, so the true cutoff is hidden even from low multiplicity derivative interactions.  
%On completely general grounds, a renormalizable UV completion cannot exhibit a mirage cutoff---by definition any such theory has low multiplicity interactions that will be  sensitive to the true cutoff.
%This simple fact may explain why field space boundaries have not be encountered previously from top-down constructions.  

Second, it would be interesting to explore the interplay between field space boundaries and symmetry.
For example, the wall potential in \Eq{V_wall} describes a scalar theory with a shift symmetry that is explicitly broken in a way that is {\it delocalized in field space}.  The breaking of the shift symmetry is not soft---indeed it is achieved by an infinite tower of higher dimension operators.  Hence, field space boundaries somehow achieve {\it hard breaking} that is radiatively stable.    It would be interesting to apply this mechanism to other symmetries.  For instance, perhaps Lorentz symmetry could be explicitly broken nonlocally on the field space boundary without inducing relevant Lorentz violating operators, in the spirit of extra-dimensional sequestering.

Finally, there is the question of whether field space boundaries can enrich any physics beyond the SM that directly utilizes large coherent fields.  Here some possible targets of study are modifications of the QCD and electroweak phase transitions, baryogenesis, Higgs vacuum instability, and inflation model building.

\medskip

\noindent {\it Acknowledgments:} 
We are grateful to Nima Arkani-Hamed, Tim Cohen, Walter Goldberger, Andreas Helset, Michele Papucci, Aneesh Manohar, Julio Parra-Martinez, Riccardo Penco, Ryan Plestid, Grant Remmen, Eva Silverstein, Mark Wise, and Kathryn Zurek for insightful discussions.
C.C.~is supported by the Department of Energy (Grant No.~DE-SC0011632) and by the Walter Burke Institute for Theoretical Physics.  I.Z.R.~is supported by the Department of Energy (Grant No.~DE-FG02-04ER41338 and FG02-06ER41449). 
%The authors would like to thanks.... Nathaniel Craig, Alberto Nicolis, 

\appendix
\section{Physical Bar}
\label{app:bar}

A great deal of insight into field space boundaries can be gleaned from the physical example of a finite solid.  This real-world system gives credence to the idea that field space endpoints are not actually all that exotic.  In fact, they are quite ordinary.

To illustrate this point, let us review the EFT description of a thin, solid bar of finite extent \cite{Nicolis:2022llw}.  This system exhibits longitudinal phonons which are effectively one-dimensional.  These degrees of freedom are described by the comoving coordinates $\phi(x,t)$, which at a time $t$ map the point $x$ in physical space to a point $\phi$ in the internal space of labeled matter elements.  
Importantly, for an object of finite volume, the domain of allowed labels $\phi$ is bounded.  We emphasize that this should not be viewed as a bound on the spacetime coordinates themselves.
%  Since we are interested 
%in the field bound we will work with the $\phi^I$ coordinates. 
%If we consider a one dimensional solid embedded, as described in \cite{},  in a higher dimensional
%space we introduce the embedding coordinate $X^\mu(\sigma,\tau)$ which maps out the world volume. Such a solid will have the action nuch like
%a fundmental string except that boost invariance along the bar  introduces an additional longitudinal degree of freedom $\phi(\sigma,\tau)$
%which labels the elements of the string by $\sigma$, and this $\phi$ is bounded by $\bar \phi$.
 %Two of the degrees of freedom in $X^\mu$ can be removed by a gauge choice ($X^0=t,X^3=\sigma$) leaving $\phi$ and two transverse modes,  
% which are unbounded and not of interest to us in this problem.

The field space boundary is implemented
by multiplying the EFT Lagrangian by a step function,
\eq{
S= \int d^4x \, \theta(\bar \phi-\phi)L(\partial \phi).
}{S_bar}
Aside from the endpoints of the bar, $\phi$ is derivatively coupled.  This makes sense because it is the Nambu-Goldstone boson of broken space translations.  Since the value of $\phi$ is free to vary at the endpoints, we have implicitly chosen open boundary conditions.  In contrast, clamping the bar would fix the value of $\phi$ at the ends.  However, the case of open boundary conditions is of particular interest because it leads to a dynamical relaxation of certain relevant operators \cite{Nicolis:2022llw}. 
%This
%infrared effect crucially depends upon the breaking of translational invariance and is thus irrelevant to the general class of models we discuss in this paper.

The EFT of a bar is valid up to a scale $\tilde \Lambda$ corresponding to the appearance of short distance modes associated with the finite  thickness of the bar.  Above this scale, this theory is UV completed into a fully three-dimensional solid which in turn has an EFT cutoff given by the atomic spacing.

The finite bar system offers some insight into the radiative stability of field space boundaries.  Since the step function in \Eq{S_bar} explicitly breaks the shift symmetry of $\phi$, it is natural to expect that radiative corrections will compound this breaking.  However, it is important to realize that the original action is still {\it exactly invariant} under the nonlinear shift symmetry $\phi \rightarrow \phi + \mu \theta(\bar\phi - \phi)$ for infinitesimal parameter $\mu$.  To preserve this symmetry, all radiatively generated terms must be localized to the ends of the bar, so they are proportional to
 $\delta(\phi-\bar \phi)$ and its derivatives. Physically, these terms encode the surface microphysics residing at the ends of the bar.   Crucially, we see that the location of the endpoint in field space---namely, the end of the bar---is radiatively stable.  
% generate the action for the surface physics. We see clearly now  the system with the bound is clearly radiatively
%stable as discussed in the body of the paper.

In a physically realistic situation, the ends of the bar will also taper off at some rate.
We can model this effect by regulating the step function to
\begin{equation}
\theta(\bar \phi -\phi)= \lim_{\Lambda \rightarrow 0}  \left(1-\tanh\left(\frac{ \bar \phi -\phi}{ \Lambda}\right)\right),
\end{equation}
where $ \Lambda$ is the tapering scale.  Note that at momenta of order $\Lambda$, new physics comes into play which is dictated by the precise shape of the ends.  In general, we expect that $ \Lambda \lesssim \tilde \Lambda$ since the
tapering length is bounded by the thickness of the bar.

\section{Balls, Springs, Strings}
\label{app:balls_springs_strings}

Even though field space boundaries appear in physical systems, it is not easy to write down an explicit UV complete Lagrangian for these dynamics.  Nevertheless, it is relatively straightforward to devise a many body system whose mean field approximation exhibits a field space boundary.   Here we present a simple example of this constructed from balls, springs, and strings.

Consider an infinite one-dimensional infinite of balls of mass $m$, each connected to its neighbor by a spring with constant $k$.  The Lagrangian is
\eq{
L = \sum_n \frac{1}{2} m \dot x_n^2 - \frac{1}{2} k (x_{n+1} - x_n -a)^2,
}{}
where $a$ is the equilibrium distance between balls.  Transforming to coordinates describing the deviations from equilibrium positions
\eq{
x_n = na + y_n,
}{}
we obtain the resulting Lagrangian
\eq{
L = \sum_n \frac{1}{2} m \dot y_n^2 - \frac{1}{2} k (y_{n+1} - y_n )^2.
}{}
At this point we add an additional force to the system.   Imagine that each ball is tethered its associated equilibrium point by a string of fixed length $\bar y$, so field space is effectively bounded by $|y_n| < \bar y$.  The Lagrangian is
\eq{
L = \sum_n \frac{1}{2} m \dot y_n^2 - \frac{1}{2} k (y_{n+1} - y_n )^2- U(y_n),
}{}
where we have defined the wall potential
\eq{
U_{\rm wall}(y_n)  &= \left\{ 
\begin{array}{ll}
0 &,\quad  |y_n|  < \bar y\\
\infty &,\quad |y_n|  >\bar y
\end{array}
\right. ,
}{}
which leashes each ball to its equilibrium point in a way that exerts no force unless the string becomes taut.

Moving to the continuum limit, we send
\eq{
y_n &\rightarrow y(x) \\
y_{n+1}- y_n &\rightarrow a \frac{\partial}{\partial x} y(x) \\
\sum_n &\rightarrow \frac{1}{a} \int dx,
}{}
so the Lagrangian becomes an integral over space,
\eq{
L = \int dx \left[ 
\frac{1}{2a} m\dot y^2 - \frac{1}{2} ka \left(\frac{\partial y}{\partial x}\right)^2 - \frac{1}{a} U_{\rm wall}(y)
\right].
}{}
Next, we rescale the field to go to our final field variable,
\eq{
\phi(x) = \sqrt{\frac{m}{a}} y(x),
}{}
and the final formula for the Lagrangian is
\eq{
L = \int dx \left[ 
\frac{1}{2} \dot \phi^2 - \frac{1}{2} v^2  \left(\frac{\partial \phi}{\partial x}\right)^2 - V_{\rm wall}(\phi)
\right],
}{}
where we have defined the velocity squared $v^2 = ka^2 /m$ and the potential $V_{\rm wall}(\phi) = U_{\rm wall}(y)/a$.

%Our EFT demands a potential $V(\phi)$ which is flat, and then turns up at the field value $\bar \phi$, forming a barrier.   For the ball and spring system, this corresponds to a force on each ball that does nothing until the ball attempts to extend more than $\bar\phi$ away from its equilibrium position.  A physical picture for this force would be a string of finite length, tethered to the ball and to the equilibrium position, that exerts no force until the ball strays too far.

\section{Unitarity Violation}
\label{app:scattering}

The precise definition of the cutoff of an EFT is given by the energy scale at which scattering amplitudes violate unitarity.  Here we work through an explicit calculation that examines this limit more carefully.

 To start, let us include a source term $J$ that couples to the boundary potential $V(\phi)$ via
\eq{
J\times  \epsilon \Lambda^4  \sum_{n=0}^\infty \frac{c_n}{n!} \left(\frac{\phi^2}{\Lambda^2} \right)^n.
}{J_source_scalar}
Here $J$ is a proxy for some incoming degrees of freedom that merge to produce a spray of $\phi$ particles, so we are interested in the $2n$-point process $J\rightarrow \phi^{2n}$ where $n \gg 1$.  %Here $n$ can either be smaller or larger than $\log(1/\epsilon)$, which from \Eq{n_crit} is the critical multiplicity at which the boundary becomes apparent.  
For example, $J$ could represent a field decaying into $\phi$, or a pair of fields annihilating into $\phi$. 
% Here we choose $p_\mu = (E, 0,0,0)$, where $E$ is the center of mass energy of the process.  

%By the optical theorem, the cross-section for this process is proportional to the discontinuity ${\rm Im} \langle J(p) J(-p) \rangle$.

The rate for the $2n$-point process is proportional to
\eq{
R(J\rightarrow \phi^{2n}) = \frac{\kappa }{(2n)!} \int d \Phi_{2n} |A_{2n}|^2,
}{}
where $\kappa$ is a constant and $d\Phi_{2n}$ is the phase space density for the final state of $2n$ particles,
\eq{
d\Phi_{2n} = %(2\pi^4) 
\delta^4\left(p - \sum_{i=1}^{2n} p_i\right) \prod_{i=1}^{2n} d^4 p_i \delta(p_i^2) \theta(p_i^0),%\frac{d^3p_i}{(2\pi)^3 2 E_i}
}{}
 and the corresponding scattering amplitude is
 \eq{
 A_{2n} &= \frac{(2n)!}{n!}\frac{\epsilon c_n }{\Lambda^{2n-4}}.
 }{}
  Since the amplitude is constant, we can use the explicit formula for the volume of phase space \cite{Kleiss:1985gy},
\eq{
 \int d \Phi_{2n}  &= \left(\frac{\pi}{2}\right)^{2n-1} \frac{E^{4n-4}}{(2n-1)!(2n-2)!} \sim   \left(E/2n\right)^{4n},
 % \\
 %& 
 %\sim \frac{4n^2}{\pi^2 E^4} \left(\frac{e}{n} \sqrt{\frac{\pi}{8}}E\right)^{4n},%\sim \left(\frac{E}{n}\right)^{2n},
}{dPhi_n}
where $E$ is the center-of-mass energy and the last line shows the leading scaling in the high multiplicity limit $2n\gg 1$.
% and in then dropped all factors except those that scale as a power of $n$ and that depend on $n$ or $E$.  
This formula encodes the intuition that the center-of-mass energy is finely partitioned into an approximate energy per particle $E/2n$,
%average energy for each leg,
%\eq{
%E_{\rm avg} &=e \sqrt{\frac{\pi}{2}} \frac{E}{n},
%}{}
so phase space contracts accordingly at high multiplicity.
Putting everything together, we obtain 
\eq{
R(J\rightarrow \phi^{2n}) &=\frac{4\kappa \epsilon^2\Lambda^8 }{\pi E^4}\frac{}{} \frac{c_n^2}{n! (n-1)! (2n-2)!} \left( \frac{\pi^{1/2}}{4^{1/4}}\frac{E}{\Lambda}\right)^{4n}\\
 &\sim  \left( \frac{ E/2n}{\epsilon^{-1/2n} \Lambda }\right)^{2n}\sim   \left( \frac{ E/2n}{\Lambda_n }\right)^{2n},
% &\overset{n\gg 1}{=} 
%\frac{\kappa \epsilon^2 c_n^2  n^2 \Lambda^8}{\pi^2 E^4} \left(\frac{e^{k+1/2}}{n^{k+1/2}} \sqrt{\frac{\pi}{2}}\frac{  E}{\Lambda}\right)^{2n}\\
%\epsilon^2 \left( \frac{ E}{\Lambda n}\right)^{2n}
% &\sim  \left( \frac{ E/n}{\Lambda_n }\right)^{2n},
}{rate_exclusive}
where we have defined the mirage cutoff,
\eq{
\Lambda_n \sim \epsilon^{-1/2n } \Lambda ,
}{}
which is in agreement with \Eq{Lambda_n}.

Let us unpack the physical significance of \Eq{rate_exclusive}.
To maintain unitarity at $2n\gg 1$, the quantity in parentheses in \Eq{rate_exclusive} must be smaller than one, so the energy per particle must be below the mirage cutoff, so 
\eq{
E/2n \lesssim \Lambda_n.
}{E_avg_bound}
From the $n$ dependence in $\Lambda_n$ we see immediately that the dynamics depend sensitively on whether we consider low or high multiplicity processes, corresponding to $n\ll n_*$ and $n\gg n_*$, respectively, where 
%$n_*$ is defined in \Eq{n_crit}.  
\eq{
n_* \sim \tfrac12 \log(1/\epsilon) \sim \tfrac12 \bar\phi^2/\Lambda^2 \gg 1,
%  \log(1/\epsilon)= \bar \phi/\Lambda,
%\lim_{n\rightarrow\infty} \Lambda_n = \Lambda,
}{}
and the induced boundary in field space is at
\eq{
\bar \phi \sim \Lambda \sqrt{\log (1/\epsilon)}.
%\epsilon = e^{-k(\bar\phi/\Lambda)^{1/k} } .
}{}
At low multiplicities, the perceived cutoff is parametrically higher than $\Lambda$, which is the phenomenon of the mirage cutoff.  Meanwhile, at high multiplicities the mirage dispels since $\Lambda_n \sim \Lambda$ and the true cutoff emerges.  In this regime the energy per particle is bounded by $E/2n \lesssim \Lambda$ to remain within the regime of validity of the EFT.
%.   In the latter case we find that
%\eq{
%E/2n \lesssim \Lambda.
%}{E_avg_bound}
In conclusion, we learn that $\Lambda$ is the physical cutoff of the EFT, serving as a proxy for the energy scale of new physics that enters to unitarize scattering.  For this reason, all divergent loop integrals in the EFT should be regulated by $\Lambda$ as well.  
%Given our starting potential in \cite{Hook:2023pba,Chang:2019vez}

Note that the unitarity bound in \Eq{E_avg_bound} also implies that the center-of-mass energy cannot be arbitrarily large within the EFT.    In particular, we find that 
\eq{
E\lesssim 2n \Lambda_n \sim \Lambda \log(1/\epsilon) \sim \bar \phi^2/\Lambda,
}{}
where we have minimized the right-hand side of the inequality on $n$ and then plugged in \Eq{barphi}.

The true cutoff $\Lambda$ is only visible in high multiplicity processes.  One might then worry
that perturbation theory---which is an asymptotic expansion---could break before attaining sufficiently high multiplicity.  However, it should be noted that the degeneration of the asymptotic series occurs when there is a proliferation of Feynman diagrams contributing to $n$-point scattering.   For processes involving the boundary potential, however, each insertion of a potential interaction is accompanied by an additional factor of the spurion $\epsilon \ll 1$.  So scattering events with more than one insertion of a boundary potential interaction are dramatically suppressed, and hence the growth of Feynman diagrams is ameliorated.

% However, since the multiplicity-dependent cutoff $\Lambda_n$ in \Eq{Lambda_n} is exponentially sensitive to the multiplicity $n$, this number needs to be sizable, but not exceedingly large to see the true cutoff.

\section{Power Law Divergent Corrections}
\label{app:power}

We have argued that the general structure of the potential in \Eq{V_gen} is radiatively stable. Here we demonstrate this in an explicit example by computing the {\it power law divergent} corrections to the exponential potential,
\eq{
V_{\rm exp} (\phi) %&= \epsilon \times (\textrm{entire function of } \phi/\Lambda) \\ 
&= \epsilon \Lambda^4  \sum_{n=0}^\infty \frac{1}{n!} \left(\frac{\phi^2}{\Lambda^2} \right)^n ,
}{V_exp2}
working to all loop orders.  Of course, power divergences are famously scheme dependent---for example, in dimensional regularization they vanish entirely.  For our analysis here we will regulate quadratic divergences as $\Lambda^2/16\pi^2$, where $\Lambda$ is a proxy for the scale at which unitarity is violated and new physics must enter. 
As we will see, in this scheme the power law divergent loop corrections renormalize the numerical coefficients of the potential while maintaining its form.

Since $\epsilon$ is small, it suffices to compute to leading order in this parameter.  The  radiative correction to the $2k$-point vertex from the $2n$-point vertex with $2(n-k)$ legs closed up into an $(n-k)$-loop diagram is
\eq{
\left[ \frac{\phi^{2k}}{(2k)!} \right] \left[ \frac{(2n)!}{n!} \frac{\epsilon }{\Lambda^{2n-4}} \right] \left[\frac{ \Lambda^2}{16\pi^2} \right]^{n-k} \left[ \frac{1}{2^{n-k}(n-k)!}\right],
}{k_to_n_potential}
where every loop is a quadratically divergent tadpole.
Of the terms in square brackets,  the first is the operator corresponding to the $2k$-point interaction, the second is the $2n$-point vertex, the third is the product of quadratically divergent loop integrals each taken to be $\Lambda^2/16\pi^2$, and the fourth  is the symmetry factor of the $(n-k)$-loop diagram.  Summing \Eq{k_to_n_potential} for  $n=k,\ldots, \infty$, we obtain a closed formula for the exponential potential corrected by power law divergences at all loop orders,
\eq{
V_{\textrm{exp, div}}(\phi)  &= \epsilon \Lambda^4  \sum_{n=0}^\infty \left(1-\frac{1}{8\pi^2}\right)^{-n-1/2}  \frac{1}{n!} \left(\frac{\phi^2}{\Lambda^2} \right)^n ,
}{}
which agrees with \Eq{V_div}.
Note that the factor of $1/8\pi^2$ is scheme dependent because it is proportional to the squared ratio of the cutoff that regulates quadratic divergences divided by the cutoff that controls the higher dimension operators in the potential.  For simplicity we have chosen both of these cutoffs to be exactly $\Lambda$. This is the scale of unitarity violation induced by the latter, which in turn fixes the scale of new physics that defines the former.
In any case, since the renormalized potential has the same form as \Eq{V_exp2}, the field space boundary is radiatively stable.

\section{Electromagnetic Boundaries}
\label{app:LF}

We asserted in \Sec{sec:BSM} that a boundary for the electromagnetic field can be spontaneously generated by a Lagrangian of the schematic form,
\eq{
L (F) &= \epsilon  \Lambda^4  \sum_{n=0}^\infty \frac{c_n}{n!} \left(\frac{F^2}{\Lambda^4}\right)^n.
% \, {\rm tr} \left(\frac{F_{\mu\nu}}{\Lambda^2}\right)^n,
}{LF2}
%where again we note that the Lorentz indices in the field strengths can be contracted in all possible ways. 
Here we spell out an explicit construction.

To start, we assume that the Lagrangian is some unknown function $L(\sigma)$ of the scalar invariant 
\eq{
\sigma = 
\frac{F^2}{\Lambda^4} = \frac{2}{\Lambda^4} (\boldsymbol B^2 - \boldsymbol E^2).
}{}
Since this EFT is defined in terms of a Lagrangian rather than a potential, it is not guaranteed that the Hamiltonian is bounded from below, as required by consistency.    The Hamiltonian for this general theory is,
\eq{
H &= \boldsymbol E\cdot \frac{ \partial L}{\partial \boldsymbol E } - L = -\frac{4 \boldsymbol E^2 }{\Lambda^4} L'(\sigma) - L(\sigma),
}{}
after dropping total derivatives and terms proportional to the Gauss law constraint.
A sufficient condition for a Hamiltonian that is bounded below is that $L(\sigma)\leq 0$ and $L'(\sigma) \leq 0$.  These conditions are satisfied for any Lagrangian that is negative and monotonically decreasing, for example like 
$L(\sigma) \sim - \exp\left(\sigma\right)$.   This class of functions includes the electromagnetic analog of \Eq{V_exp},
\eq{
L(F)  = -\Lambda^4 \exp\left(\frac{F^2- \bar F^2}{\Lambda^4}\right),
}{}
where $\bar F$ is the boundary of the electromagnetic field. Like before, we can series expand the Lagrangian,
\eq{
L(F)= - \epsilon \Lambda^4  \sum_{n=0}^\infty \frac{1}{n!} \left(\frac{F^2}{\Lambda^4} \right)^n , %=  -  \sum_{n=0}^\infty \frac{1}{n!} \frac{F^{2n}}{\Lambda_n^{4n-4}} ,
}{}
where $\epsilon = \exp(-\bar F^2/\Lambda^4)\ll1 $ is a tiny spurion.
% and the mirage cutoff is
%\eq{
%\Lambda_n = \Lambda \epsilon^{-1/(4n-4)}.
%}{}
%For $2n$-point scattering in a trivial background, the true cutoff is invisible until the critical multiplicity
%\eq{
%2n%-4
%\gg \tfrac12 \log(1/\epsilon)= \tfrac12  \bar  F^2/\Lambda^4,
%\lim_{n\rightarrow\infty} \Lambda_n = \Lambda,
%}{n_crit_EM}
%which signals the appearance of the field space boundary.

Next, let us determine the regime of validity of this EFT.  In analogy with \Eq{J_source_scalar}, we couple a source $J$ to the electromagnetic field boundary,
\eq{
-J\times \epsilon  \Lambda^4  \sum_{n=0}^\infty \frac{c_n}{n!} \left(\frac{F^2}{\Lambda^4}\right)^n,
}{}
and consider the $2n$-point process $J\rightarrow \gamma^{2n}$.  While the associated rate is too cumbersome to calculate exactly, it can be easily estimated.  
Approximating the phase space integral by \Eq{dPhi_n} and the scattering amplitude by  
\eq{
A_{2n} \sim \frac{ (2n)!}{n!} \frac{\epsilon c_n}{\Lambda^{4n-4}} \left(\frac{E}{2n}\right)^{2n} ,%\\ \epsilon(E/2n)^{2n} /\Lambda^{4n-4},
}{}
 we obtain the rate
\eq{
R(J\rightarrow \gamma^{2n})& \sim \left(\frac{  E/2n}{\epsilon^{-1/4n}\Lambda }\right)^{8n} \sim  \left( \frac{ E/2n}{\Lambda_n }\right)^{8n},
}{}
where we have defined the mirage cutoff for the electromagnetic field to be
\eq{
\Lambda_n \sim \epsilon^{-1/4n } \Lambda .
}{}
Like before, we find that unitarity requires that the energy per particle be less than the mirage cutoff, so \eq{
E/2n \lesssim \Lambda_n.
}{}
Again, the $n$ dependence in $\Lambda_n$ defines a critical multiplicity for which $n$ exceeds the value
\eq{
n_* \sim \tfrac14 \log(1/\epsilon) \sim \tfrac14 \bar F^2/\Lambda^4.
}{}
Last not least, the remain within the regime of validity of the EFT, the center-of-mass energy is bounded by
\eq{
E  \lesssim 2n \Lambda \sim \tfrac12 \Lambda \log(1/\epsilon) \sim \tfrac12  \bar F^2/\Lambda^3,
}{}
where we have again minimized the right-hand side of the inequality as a function of $n$.
%Thus, in the regime of validity of the EFT, the center-of-mass energy must not exceed a particular scale that depends on the value of the electromagnetic field boundary and the true cutoff.

\bibliographystyle{utphys-modified}
\bibliography{bar}

\providecommand{\href}[2]{#2}\begingroup\raggedright\begin{thebibliography}{10}

\bibitem{Note1}
This statement applies to the potential, but in general one must also grapple
  with low multiplicity derivative interactions that are set by the scale
  $\Lambda $ without any additional suppression by $\epsilon $. In this case
  the true cutoff $\Lambda $ is accessible at low multiplicity and there is no
  mirage. For example, unsuppressed interactions of this kind appear in many UV
  completions of the axion after integrating out an accompanying radial Higgs
  mode or tower of extra-dimensional excitations. Still, it is still striking
  that the axion potential exhibits a mirage cutoff.

\bibitem{Hook:2023pba}
A.~Hook and R.~Rattazzi, ``{Softening the UV without new particles},''
  \href{http://dx.doi.org/10.1103/PhysRevD.108.115019}{{\em Phys. Rev. D}
  {\bfseries 108} no.~11, (2023) 115019},
  \href{http://arxiv.org/abs/2306.12489}{{\ttfamily arXiv:2306.12489
  [hep-ph]}}.

\bibitem{Note2}
Alternatively, we can enforce a boundary on field space by multiplying the
  kinetic energy term by a step function $\theta (\protect \bar \phi -\phi )$.
  The physical equivalence between an infinite wall in the potential energy
  versus a step function in the kinetic energy is analogous to what happens in
  Einstein versus Jordan frame in gravity. For a detailed review of the EFT of
  phonons in a physical bar, see App.~\ref {app:bar}. In App.~\ref
  {app:balls_springs_strings} we also present a simple model of balls, springs,
  and strings whose mean field approximation exhibits a boundary in field
  space.

\bibitem{Note3}
Notably, boundaries in {\protect \it coordinate space} have long appeared in
  the context of gravity as geometric cutoffs or ``Dirichlet walls''. Since
  these boundaries are not dynamical branes, they need not be stabilized, but
  at the same time their consistency is not guaranteed. Nevertheless, under
  certain conditions, these endpoint objects have been found to be stable \cite
  {Andrade:2015gja,Andrade:2015qea,Marolf:2012dr} and consistent with the
  initial value problem \cite {An:2021fcq}.

\bibitem{Note4}
Here we define our potential to be a series expansion in $(\phi ^2/\Lambda
  ^2)^n$ with coefficients that scale as $1/n!$, as opposed to a series
  expansion in $(\phi /\Lambda )^n$ with coefficients that scale as $1/n!$. We
  do this for two reasons. Firstly, our phenomenological applications all
  involve field space boundaries for charged fields like the Higgs boson or the
  SM force carriers, which only appear in even powers in their
  self-interactions. The second reason has to do with a minor subtlety
  regarding the scale of unitarity violation. A general potential defined by a
  series expansion in $(\phi /\Lambda )^n$ with coefficients $1/n!$ will cover
  the familiar case of a periodic scalar, as well as the exponential runaway of
  a dilaton field. However, as shown in \cite {Hook:2023pba}, these models
  exhibit scattering which at {\protect \it any} multiplicity will never see a
  breakdown of the EFT at the scale $\Lambda $ but rather at the slightly
  higher cutoff $\Lambda \protect \sqrt {\log (1/\epsilon )}$, which can be
  realized in explicit UV models \cite {Hook:2023pba}. Of course, these factors
  of $\log (1/\epsilon )$ corrections are far less important than the powers of
  $\epsilon $ that generate the phenomenon of the mirage cutoff. Nevertheless,
  for simplicity we restrict here to a potential which is a series expansion in
  $(\phi ^2/\Lambda ^2)^n$ with coefficients $1/n!$, which as shown in
  App.~\ref {app:scattering} has a physical cutoff defined by unitarity
  violation in scattering that is precisely the scale $\Lambda $, with no
  additional factors of $\log (1/\epsilon )$. Note that in general, changing
  the power of $\phi $ in the series expansion has the same effect as changing
  the scaling of the series coefficient. So for example, a series expansion in
  $(\phi ^k/\Lambda ^k)^n$ with coefficients $1/n!$ is the same as a series
  expansion in $(\phi /\Lambda )^n$ with coefficients $(1/n!)^{1/k}$.

\bibitem{Note5}
It may be concerning that this potential---like that of the axion---is an
  entire function. In particular, there is evidence that the generating
  functional of connected correlators $W[J]$ cannot be entire \cite
  {Panagopoulos:2020sxp,Cohen:2022clv}, which is related to the asymptotic
  nature of the perturbative series \cite
  {Brown:1992ay,Son:1995wz,Rubakov:1995hq,Libanov:1997nt}. Furthermore, in
  certain explicit examples the operator product expansion can also be
  asymptotic \cite {Dunne:1999uy}. Nevertheless, we note that these
  degenerations at high multiplicity all stem from the growth of diagrammatic
  topologies in {\protect \it correlators}, while the potential defines
  individual {\protect \it interaction vertices}. As shown in App.~\ref
  {app:scattering}, any contribution to a correlators involving multiple
  insertions of the potential interactions will be subleading since $\epsilon
  \ll 1$. Consequently, the growth in diagrams is tamed and any subtleties
  concerning the asymptotic nature of perturbation theory are not relevant.

\bibitem{Note6}
As shown in Sec.~\ref {sec:naturalness}, radiative corrections serve only to
  renormalize the coefficients $c_n$ to numerical values of the same order.
  Since this manifestly preserves the structure of the potential in Eq.~(\ref
  {V_gen}), it also maintains the flatness of the potential for $|\phi |
  \lesssim \protect \bar \phi $. Nevertheless, these corrections could in
  principle still flip the signs of $c_n$ in such a way that destabilizes the
  potential, toggling the feature at $|\phi | \sim \protect \bar \phi $ from a
  wall into a cliff. This is reminiscent of the sensitivity of vacuum alignment
  to quantum corrections, for example as occurs in radiative electroweak
  symmetry breaking. Here also, the binary question of whether the potential
  develops a wall or a cliff will be decided by the UV dynamics that regulate
  the divergences, so in this sense these two possibilities have even odds.
  Notably, if the underlying theory is supersymmetric, then the potential
  energy will be automatically bounded from below.

\bibitem{Note7}
One should bear in mind that the EFT transitions from the weak field to strong
  field regime long before $\phi $ reaches $\protect \bar \phi $. To see why,
  recall that the weak field regime by definition applies when perturbation
  theory in the field value is accurate. This implies that the $n$-point
  interaction in $V(\phi )$ must be more important than the $(n+1)$-point
  interaction. From $V(\phi )$ it is clear that this transition occurs at the
  critical field value $|\phi | \sim \Lambda \ll \protect \bar \phi $.

\bibitem{Note8}
The potential terms generate derivative interactions in the same way as in
  $\phi ^4$ theory, where wavefunction renormalization appears at two loops.

\bibitem{Note9}
What is the natural size for these derivatively coupled interactions? Consider,
  for example, the $ (\partial \phi )^4$ operator, whose coefficient is
  famously constrained to be nonnegative by analyticity, unitarity, and
  causality \cite {Adams:2006sv}. In particular, this coefficient is equal to
  an energy weighted integral over the total cross-section for $\phi \phi
  \rightarrow \chi $, where $\chi $ denotes the heavy states in the UV
  completion. To preserve the phenomenon of the mirage cutoff, the coefficient
  of $ (\partial \phi )^4$ must be suppressed by $ \epsilon \ll 1$, which
  implies that $\chi $ must couple very weakly to a pair of $\phi $ particles.
  This is very different than what happens in traditional UV completions, like
  the linear sigma model completion of a periodic scalar. Conversely, an EFT
  with a mirage cutoff can only arise in a UV completion in which the heavy
  state $\chi $ only couples to a {\protect \it high multiplicity} of $\phi $
  particles.

\bibitem{Note10}
If the mass of $\phi $ is calculable as a function of the parameters of the UV
  theory, then the dependence on $\protect \tilde \Lambda $ will emerge from
  the physical mass scale of the heavy degrees of freedom. This happens, for
  example, in phenomenological models of soft supersymmetry breaking. If the
  mass of $\phi $ is incalculable, however, then its quadratic divergences must
  cancelled by a counterterm whose value is uniquely fixed by the physical
  mass, which can only be extracted from experimental observations. In this
  case, the mass cannot be predicted in terms of UV parameters, and technically
  there is no hierarchy problem.

\bibitem{ParticleDataGroup:2024cfk}
{\bfseries Particle Data Group} {\bfseries Collaboration}, S.~Navas { et~al.},
  ``{Review of particle physics},''
  \href{http://dx.doi.org/10.1103/PhysRevD.110.030001}{{\em Phys. Rev. D}
  {\bfseries 110} no.~3, (2024) 030001}.

\bibitem{ATLAS:2012yve}
{\bfseries ATLAS} {\bfseries Collaboration}, G.~Aad { et~al.}, ``{Observation
  of a new particle in the search for the Standard Model Higgs boson with the
  ATLAS detector at the LHC},''
  \href{http://dx.doi.org/10.1016/j.physletb.2012.08.020}{{\em Phys. Lett. B}
  {\bfseries 716} (2012) 1--29},
  \href{http://arxiv.org/abs/1207.7214}{{\ttfamily arXiv:1207.7214 [hep-ex]}}.

\bibitem{CMS:2012qbp}
{\bfseries CMS} {\bfseries Collaboration}, S.~Chatrchyan { et~al.},
  ``{Observation of a New Boson at a Mass of 125 GeV with the CMS Experiment at
  the LHC},'' \href{http://dx.doi.org/10.1016/j.physletb.2012.08.021}{{\em
  Phys. Lett. B} {\bfseries 716} (2012) 30--61},
  \href{http://arxiv.org/abs/1207.7235}{{\ttfamily arXiv:1207.7235 [hep-ex]}}.

\bibitem{Alonso:2015fsp}
R.~Alonso, E.~E. Jenkins, and A.~V. Manohar, ``{A Geometric Formulation of
  Higgs Effective Field Theory: Measuring the Curvature of Scalar Field
  Space},'' \href{http://dx.doi.org/10.1016/j.physletb.2016.01.041}{{\em Phys.
  Lett. B} {\bfseries 754} (2016) 335--342},
  \href{http://arxiv.org/abs/1511.00724}{{\ttfamily arXiv:1511.00724
  [hep-ph]}}.

\bibitem{Note11}
Is there a version of electroweak symmetry breaking that directly utilizes a
  boundary in Higgs field space? The simplest version of this idea is to set
  $\lambda =0$, so the tachyonic mass for the Higgs field drives its VEV to the
  boundary of field space. For $\Lambda \ll \protect \bar H$, which is required
  for radiative stability of the boundary, the VEV is pushed to the edge,
  $\langle H\rangle \sim \protect \bar H$, as expected, irrespective of the
  value of tachyonic mass $m^2$. Unfortunately, by expanding in fluctuations,
  $H = \langle H\rangle + h$, we see that the Higgs particle exhibits nonlinear
  interactions that enter in powers of $h \protect \bar H/ \Lambda ^2$, which
  are enhanced for $\Lambda \ll \protect \bar H$. Since the Higgs VEV is pinned
  inside the cusp between the tachyonic mass and the boundary, the Higgs
  particle is strongly coupled and thus not viable.

\bibitem{Arkani-Hamed:1998jmv}
N.~Arkani-Hamed, S.~Dimopoulos, and G.~R. Dvali, ``{The Hierarchy problem and
  new dimensions at a millimeter},''
  \href{http://dx.doi.org/10.1016/S0370-2693(98)00466-3}{{\em Phys. Lett. B}
  {\bfseries 429} (1998) 263--272},
  \href{http://arxiv.org/abs/hep-ph/9803315}{{\ttfamily arXiv:hep-ph/9803315}}.

\bibitem{Note12}
We have imposed a boundary on the electromagnetic field so as to preserve gauge
  invariance. In principle, we could instead consider an infinite wall boundary
  on the gauge field itself, corresponding to a Proca potential which has been
  sequestered to high multiplicity. If we regulate the boundary wall to have
  finite thickness, then tiny but nonzero Proca potential terms will be induced
  in the trivial vacuum. While such a setup might be viable for an abelian
  gauge field like hypercharge, this is likely pathological for the nonabelian
  case, where Proca mass terms invariably lower the cutoff.

\bibitem{Note13}
Strictly speaking, the number of photons produced in a given scattering process
  is ill-defined on account of infrared divergences. However, since the field
  strength associated with a soft photon is vanishingly small, it essentially
  decouples from all higher dimension operators. This accords with the
  calculation in App.~\ref {app:LF}, which shows that EFT breaks down when the
  energy per particle exceeds the cutoff.

\bibitem{Ellis:2017edi}
J.~Ellis, N.~E. Mavromatos, and T.~You, ``{Light-by-Light Scattering Constraint
  on Born-Infeld Theory},''
  \href{http://dx.doi.org/10.1103/PhysRevLett.118.261802}{{\em Phys. Rev.
  Lett.} {\bfseries 118} no.~26, (2017) 261802},
  \href{http://arxiv.org/abs/1703.08450}{{\ttfamily arXiv:1703.08450
  [hep-ph]}}.

\bibitem{ATLAS:2017fur}
{\bfseries ATLAS} {\bfseries Collaboration}, M.~Aaboud { et~al.}, ``{Evidence
  for light-by-light scattering in heavy-ion collisions with the ATLAS detector
  at the LHC},'' \href{http://dx.doi.org/10.1038/nphys4208}{{\em Nature Phys.}
  {\bfseries 13} no.~9, (2017) 852--858},
  \href{http://arxiv.org/abs/1702.01625}{{\ttfamily arXiv:1702.01625
  [hep-ex]}}.

\bibitem{dEnterria:2013zqi}
D.~d'Enterria and G.~G. da~Silveira, ``{Observing light-by-light scattering at
  the Large Hadron Collider},''
  \href{http://dx.doi.org/10.1103/PhysRevLett.111.080405}{{\em Phys. Rev.
  Lett.} {\bfseries 111} (2013) 080405},
  \href{http://arxiv.org/abs/1305.7142}{{\ttfamily arXiv:1305.7142 [hep-ph]}}.
  [Erratum: Phys.Rev.Lett. 116, 129901 (2016)].

\bibitem{Knapen:2016hky}
S.~Knapen, S.~Pagan~Griso, M.~Papucci, and D.~J. Robinson, ``{Triggering Soft
  Bombs at the LHC},'' \href{http://dx.doi.org/10.1007/JHEP08(2017)076}{{\em
  JHEP} {\bfseries 08} (2017) 076},
  \href{http://arxiv.org/abs/1612.00850}{{\ttfamily arXiv:1612.00850
  [hep-ph]}}.

\bibitem{CMS:2024nca}
{\bfseries CMS} {\bfseries Collaboration}, A.~Hayrapetyan { et~al.}, ``{Search
  for soft unclustered energy patterns in proton-proton collisions at 13
  TeV},'' \href{http://arxiv.org/abs/2403.05311}{{\ttfamily arXiv:2403.05311
  [hep-ex]}}.

\bibitem{Abramowicz:2019gvx}
H.~Abramowicz { et~al.}, ``{Letter of Intent for the LUXE Experiment},''
  \href{http://arxiv.org/abs/1909.00860}{{\ttfamily arXiv:1909.00860
  [physics.ins-det]}}.

\bibitem{Panagopoulos:2019ail}
G.~Panagopoulos and E.~Silverstein, ``{Primordial Black Holes from non-Gaussian
  tails},'' \href{http://arxiv.org/abs/1906.02827}{{\ttfamily arXiv:1906.02827
  [hep-th]}}.

\bibitem{Panagopoulos:2020sxp}
G.~Panagopoulos and E.~Silverstein, ``{Multipoint correlators in multifield
  cosmology},'' \href{http://arxiv.org/abs/2003.05883}{{\ttfamily
  arXiv:2003.05883 [hep-th]}}.

\bibitem{DeLuca:2021pej}
G.~B. De~Luca, E.~Silverstein, and G.~Torroba, ``{Hyperbolic compactification
  of M-theory and de Sitter quantum gravity},''
  \href{http://dx.doi.org/10.21468/SciPostPhys.12.3.083}{{\em SciPost Phys.}
  {\bfseries 12} no.~3, (2022) 083},
  \href{http://arxiv.org/abs/2104.13380}{{\ttfamily arXiv:2104.13380
  [hep-th]}}.

\bibitem{Nicolis:2022llw}
A.~Nicolis and I.~Z. Rothstein, ``{Apparent fine tunings for field theories
  with broken space-time symmetries},''
  \href{http://dx.doi.org/10.21468/SciPostPhys.16.2.045}{{\em SciPost Phys.}
  {\bfseries 16} no.~2, (2024) 045},
  \href{http://arxiv.org/abs/2212.08976}{{\ttfamily arXiv:2212.08976
  [hep-th]}}.

\bibitem{Kleiss:1985gy}
R.~Kleiss, W.~J. Stirling, and S.~D. Ellis, ``{A New Monte Carlo Treatment of
  Multiparticle Phase Space at High-energies},''
  \href{http://dx.doi.org/10.1016/0010-4655(86)90119-0}{{\em Comput. Phys.
  Commun.} {\bfseries 40} (1986) 359}.

\bibitem{Andrade:2015gja}
T.~Andrade, W.~R. Kelly, D.~Marolf, and J.~E. Santos, ``{On the stability of
  gravity with Dirichlet walls},''
  \href{http://dx.doi.org/10.1088/0264-9381/32/23/235006}{{\em Class. Quant.
  Grav.} {\bfseries 32} no.~23, (2015) 235006},
  \href{http://arxiv.org/abs/1504.07580}{{\ttfamily arXiv:1504.07580 [gr-qc]}}.

\bibitem{Andrade:2015qea}
T.~Andrade, W.~R. Kelly, and D.~Marolf, ``{Einstein\textendash{}Maxwell
  Dirichlet walls, negative kinetic energies, and the adiabatic approximation
  for extreme black holes},''
  \href{http://dx.doi.org/10.1088/0264-9381/32/19/195017}{{\em Class. Quant.
  Grav.} {\bfseries 32} no.~19, (2015) 195017},
  \href{http://arxiv.org/abs/1503.03915}{{\ttfamily arXiv:1503.03915 [gr-qc]}}.

\bibitem{Marolf:2012dr}
D.~Marolf and M.~Rangamani, ``{Causality and the AdS Dirichlet problem},''
  \href{http://dx.doi.org/10.1007/JHEP04(2012)035}{{\em JHEP} {\bfseries 04}
  (2012) 035}, \href{http://arxiv.org/abs/1201.1233}{{\ttfamily arXiv:1201.1233
  [hep-th]}}.

\bibitem{An:2021fcq}
Z.~An and M.~T. Anderson, ``{The initial boundary value problem and quasi-local
  Hamiltonians in General Relativity},''
  \href{http://arxiv.org/abs/2103.15673}{{\ttfamily arXiv:2103.15673 [gr-qc]}}.

\bibitem{Cohen:2022clv}
T.~Cohen, D.~Green, and A.~Premkumar, ``{Large deviations in the early
  Universe},'' \href{http://dx.doi.org/10.1103/PhysRevD.107.083501}{{\em Phys.
  Rev. D} {\bfseries 107} no.~8, (2023) 083501},
  \href{http://arxiv.org/abs/2212.02535}{{\ttfamily arXiv:2212.02535
  [hep-th]}}.

\bibitem{Brown:1992ay}
L.~S. Brown, ``{Summing tree graphs at threshold},''
  \href{http://dx.doi.org/10.1103/PhysRevD.46.R4125}{{\em Phys. Rev. D}
  {\bfseries 46} (1992) R4125--R4127},
  \href{http://arxiv.org/abs/hep-ph/9209203}{{\ttfamily arXiv:hep-ph/9209203}}.

\bibitem{Son:1995wz}
D.~T. Son, ``{Semiclassical approach for multiparticle production in scalar
  theories},'' \href{http://dx.doi.org/10.1016/0550-3213(96)00386-0}{{\em Nucl.
  Phys. B} {\bfseries 477} (1996) 378--406},
  \href{http://arxiv.org/abs/hep-ph/9505338}{{\ttfamily arXiv:hep-ph/9505338}}.

\bibitem{Rubakov:1995hq}
V.~A. Rubakov, ``{Nonperturbative aspects of multiparticle production},'' in
  {\em {2nd Rencontres du Vietnam}: {Consisting of 2 parallel conferences:
  Astrophysics Meeting: From the Sun and Beyond / Particle Physics Meeting:
  Physics at the Frontiers of the Standard Model}}.
\newblock 10, 1995.
\newblock \href{http://arxiv.org/abs/hep-ph/9511236}{{\ttfamily
  arXiv:hep-ph/9511236}}.

\bibitem{Libanov:1997nt}
M.~V. Libanov, V.~A. Rubakov, and S.~V. Troitsky, ``{Multiparticle processes
  and semiclassical analysis in bosonic field theories},''
  \href{http://dx.doi.org/10.1134/1.953038}{{\em Phys. Part. Nucl.} {\bfseries
  28} (1997) 217--240}.

\bibitem{Dunne:1999uy}
G.~V. Dunne and T.~M. Hall, ``{Borel summation of the derivative expansion and
  effective actions},''
  \href{http://dx.doi.org/10.1103/PhysRevD.60.065002}{{\em Phys. Rev. D}
  {\bfseries 60} (1999) 065002},
  \href{http://arxiv.org/abs/hep-th/9902064}{{\ttfamily arXiv:hep-th/9902064}}.

\bibitem{Adams:2006sv}
A.~Adams, N.~Arkani-Hamed, S.~Dubovsky, A.~Nicolis, and R.~Rattazzi,
  ``{Causality, analyticity and an IR obstruction to UV completion},''
  \href{http://dx.doi.org/10.1088/1126-6708/2006/10/014}{{\em JHEP} {\bfseries
  10} (2006) 014}, \href{http://arxiv.org/abs/hep-th/0602178}{{\ttfamily
  arXiv:hep-th/0602178}}.

\end{thebibliography}\endgroup

\end{document}